\documentclass[letterpaper,12pt]{article}
\usepackage{color,amssymb,enumerate,float,amsmath}
\usepackage{tikz}
\usetikzlibrary{matrix}
\usepackage{caption}
\usepackage{subcaption}
\usepackage{cite}
\usepackage{ifpdf}
\ifpdf
	\usepackage[pdftex,unicode,implicit]{hyperref}

	\hypersetup{
  	pdftitle     = {}, 
  	pdfkeywords  = {},
  	pdfauthor    = {},
  	pdfcreator   = {pdf\LaTeXe\ with package \flqq hyperref\frqq},
  	pdfproducer  = {pdf\LaTeXe\ with package \flqq hyperref\frqq},
  	pdfpagemode  = UseNone,  
  	pdffitwindow = true,  
  	unicode      = true,
  	plainpages   = true,
  	colorlinks   = true,  
  	citecolor    = black,  
  	urlcolor     = blue, 
  	linkcolor    = black
	}

\else

  \usepackage[dvips]{graphicx}
  \usepackage[unicode,implicit]{hyperref}

\fi 

\makeatletter
\@addtoreset{equation}{section}
\makeatother

\begin{document}
	
\thispagestyle{empty}

\begin{center}
{\bf \LARGE One-Loop Quantum Corrections to the Casimir Effect for Smoothly Rough Plates in the Low-Temperature Regime}
\vspace*{15mm}

{\large Claudio B\'orquez}$^{1,a}$
{\large and Byron Droguett}$^{2,b}$
\vspace{3ex}

$^1${\it Facultad de Ingenier\'ia, Universidad San Sebasti\'an, Lago Panguipulli 1390, Puerto Montt, Chile.}

$^2${\it Department of Physics, Universidad de Antofagasta, 1240000 Antofagasta, Chile.
}
\vspace{3ex}

$^a${\tt claudio.borquez@uss.cl},
$^b${\tt byron.droguett@uantof.cl}

\hspace{.5em}

{\bf Abstract}
\begin{quotation}{\small\noindent
} 
We present a theoretical analysis of the one-loop effective potential of a self-interacting real scalar field in the presence of two parallel conducting plates with geometric roughness. The analysis is restricted to the adiabatic regime of smoothly varying surface deformations, where derivative contributions to the surface profile can be neglected. Using Wentzel-Kramers-Brillouin methods to evaluate the spectral density of the modified Laplace-Beltrami operator, together with contour integration within a $\zeta$-function regularization scheme, we derive analytical expressions for the quantum corrections to the effective potential induced by perturbative boundary roughness and finite temperature. Furthermore, within this regime, we compute explicit contributions to the Casimir energy and to the topological mass generation associated with the geometry.

\end{quotation}
\vspace{3ex}
\end{center}

\thispagestyle{empty}

\newpage


\section{Introduction}

The Casimir effect constitutes one of the most prominent predictions of quantum field theory, providing a macroscopic manifestation of vacuum fluctuations. In his original work, Hendrik B. Casimir demonstrated that the presence of two neutral, parallel conducting plates modifies the vacuum state of the electromagnetic field, resulting in an attractive force that depends on the separation distance \cite{Casimir:1948dh}. Since its theoretical formulation, this effect has been confirmed by numerous high-precision experiments \cite{Lamoreaux:1996wh, Bressi:2002fr}, which have turned the Casimir phenomenon into a fundamental laboratory for testing the properties of quantum vacuum. Several research has established that the Casimir interaction is not universal, but depends sensitively on boundary conditions, material properties, spacetime topology, temperature \cite{Mazur,Bordag:2001qi,Teo:2011kt,Fosco:2011xx,Bimonte:2012dqc,Zhao:2006rr,Beneventano:2004zd,Mota:2023cmr,Bellucci:2019ybj}, and external fields such as magnetic \cite{Beneventano:2005sd,Erdas:2013jga,Erdas:2015yac,Droguett:2025frq} or gravitational backgrounds \cite{Ford:1976fn,Nazari:2015oha,Nazari:2025bae,Muniz:2014dga,Borquez:2023cuf}. Furthermore, geometry plays a crucial role in determining the Casimir interaction: while parallel plates typically exhibit an attractive force, other configurations can lead to repulsion \cite{Boyer:1968uf} or even generate nontrivial lateral stresses \cite{Emig:2002qpp}, highlighting the intricate relationship between vacuum energy and spatial constraints. The study of the Casimir effect has also attracted significant attention in scenarios with Lorentz symmetry breaking. In conventional approaches, this breaking is typically implemented through the introduction of a preferred direction \cite{deMello:2022tuv, Cruz:2017kfo,Erdas:2020ilo,Cruz:2018bqt,Erdas:2021xvv,Droguett:2024tpe}. By contrast, in theories such as Ho\v{r}ava-Lifshitz gravity \cite{Horava:2009uw}, Lorentz symmetry is broken through an anisotropic scaling between space and time, characterized by a critical exponent, an idea originally formalized in the context of anisotropic field theories by Anselmi \cite{Anselmi:2008bq}. Representative studies within this framework can be found in \cite{Ferrari:2010dj,MoralesUlion:2015tve, daSilva:2019iwn,Erdas:2023wzy}.
This modification alters the spectral eigenvalue problem of the fields, leading to shifts in the vacuum energy density and the resulting Casimir pressure between the boundaries \cite{Cheng:2022mwd,Borquez:2023ajx}. In this context, the structure of the quantum vacuum becomes sensitive to the underlying spacetime symmetries, so that deviations from Lorentz invariance may, in principle, be probed through precision measurements of the Casimir force. Furthermore, these analyzes typically rely on specialized techniques, such as generalized $\zeta$-function regularization \cite{Kirsten:2007ev,Kirsten:2010zp}, to consistently handle divergences arising in the evaluation of the zero-point energy.

Recent studies have further extended the investigation of this phenomenon by considering self-interacting field theories within the effective action formalism \cite{Junior:2023feu,FariasJunior:2022qsp,Porfirio:2019gdy}. Modifications such as nontrivial spacetime topology, alternative boundary conditions, the inclusion of external fields, and even explicit Lorentz symmetry breaking have enabled the computation of new quantum corrections to the vacuum energy density at both one- and two-loop orders \cite{Toms:1979ij,Toms:1980sx,Junior:2025thl, Valuyan:2025vol, Farias:2024uzf, Junior:2024smu}.
As shown in \cite{Toms:1979ij,Toms:1980sx}, one notable consequence of self-interactions in the presence of nontrivial topology is that both massive and massless fields can acquire a topological mass through the renormalization procedure. In those works, two real scalar fields, one twisted and the other untwisted, interacting under periodic and antiperiodic boundary conditions, were analyzed. The results indicate that not only is the vacuum energy modified by topology and boundary conditions, but also that additional contributions emerge in the form of an effective mass of topological origin, which in turn plays a crucial role in determining vacuum stability.

Although most studies of the Casimir effect assume perfectly flat boundaries, real physical surfaces always exhibit some degree of roughness. Unlike the idealized case, surface roughness modifies the boundary plates and, consequently, the entire spectrum of the quantum fluctuations. We theoretically analyze the Casimir effect for a real self-interacting scalar field of the form $\Phi^{4}$, embedded in a $3+1$ dimensional spacetime, where the field is confined by a rough border that satisfies the Dirichlet boundary conditions. Furthermore, we analyze the contribution of temperature to the system implemented through periodic boundary conditions. Due to the complexity of the finite-temperature contributions, we focus on the low-temperature limit. The self-interactions of the theory have effects on quantum fluctuations and must be treated consistently. To this end, we employ the one-loop effective potential formalism \cite{Jackiw:1974cv,book1}, which allows us to renormalize the ultraviolet divergences and extract the relevant finite contributions to the Casimir energy. This method provides us with a systematic framework for examining modifications in the interactions of the vacuum structure in the presence of boundary conditions, using the well-known effective potential, which is made up of a classical contribution and quantum corrections. With respect to spatial geometry, the presence of roughness at the boundaries induces significant modifications in the spectrum of quantum fluctuations, such as was found in \cite{Droguett:2025frq,Borquez:2023ajx}. Roughness is treated perturbatively in an adiabatic regime in which the roughness profile varies smoothly along the plates, that is, we introduce a small deviation from flatness and expand the geometry to higher orders in roughness amplitude. The perturbative approach enables us to isolate the corrections induced by geometry, without losing analytic control over the eigenvalue problem. In this analysis, the dominant contribution arises from the local modification of the plate separation, while derivative terms associated with lateral contributions are parametrically suppressed. To obtain the spectrum of the confined modes, we apply the Wentzel-Kramers-Brillouin (WKB) approximation, which yields accurate estimates for the eigenvalues in geometries where exact solutions are generally inaccessible \cite{Jeffres:2012de}. With the approximate eigenvalues obtained through WKB analysis, we construct the one-loop effective potential and compute the renormalized Casimir energy for the interacting theory. By combining perturbative roughness treatment in the local adiabatic regime with the effective potential formalism, we determine the shape of the quantum vacuum corrections through the relationship between geometry, interactions, boundary conditions, and finite temperature.

The present paper is organized as follows. In Sect. 2, we define the geometric problem for the slowly varying rough boundaries and derive the field solutions using the WKB method. We also apply the contour integration of the $\zeta$-function to obtain the spectral function in the low-temperature limit. In Sect. 3, we execute the renormalization process and analyze the vacuum stability, the Casimir energy density, and the explicit form of the topological mass. Finally, in Sect. 4, we summarize our findings and present our concluding remarks.


\section{Effective potential for the self-interacting scalar field under a rough geometry}


In this section, we provide a brief introduction to the procedure for determining the one-loop contributions generated by a massive self-interacting real scalar field. To this end, we consider the Lorentzian action for the interacting scalar field, expanded up to the fourth order in the field. The potential includes the necessary counterterms, which must be fixed through the renormalization procedure. Therefore, the action can be written as
\begin{eqnarray}\label{action1}
    S[\Phi] =
    \int d^{4}x\sqrt{g}\left(\,
    \frac{1}{2}\Phi\,\Box\,\Phi 
    - \frac{m^{2} + \delta m}{2}\Phi^{2} 
    - \frac{g + \delta g}{4!}\Phi^{4}
    - \delta c\right)\,,
\end{eqnarray}
where $m$ is the mass of the field $\Phi$, $g$ is the self-interaction coupling constant, and $\delta m,\,\delta g$ and $\delta c$ are the renormalization constants. To determine the one-loop corrections generated by the one-particle-irreducible (1PI) diagrams, we construct the effective action, $\Gamma$, within the loop-expansion framework \cite{book1}. The effective action incorporates both the classical dynamics and the quantum corrections arising from fluctuations around a fixed background field, defined as the vacuum expectation value of the quantum field in the presence of external sources. The loop expansion has the form
\begin{equation}
    \Gamma[\Psi] 
    = S[\Psi]
    + \sum_{i=1}\hbar^i\, \Gamma^{(i)}[\Psi]\,,
\end{equation}
where $S[\Phi]$ is the classical action and $\Gamma^{(i)}$ denotes the contribution to the $i$-loop. The background field is introduced through the decomposition $\Phi = \Psi + \varphi$, where $\varphi$ corresponds to quantum fluctuations. Expanding the action up to the second order in $\varphi$ leads to
\begin{equation}
    S[\Psi+\varphi] = 
    S_{1}[\Psi] 
    + \frac{1}{2}\, 
    \varphi\, S_{2}[\Psi]\, \varphi
    + \mathcal{O}(\varphi^3),
\end{equation}
where $S_{2}[\Psi]$ is the second functional derivative of the classical action. All one-loop contributions arise from this quadratic operator. Performing the Gaussian functional integral,
\begin{equation}
    \exp\left( i\Gamma^{(1)}[\Psi] \right) = 
    \int \mathcal{D}\varphi\,
    \exp\!\left( \frac{i}{2}\,\int d^4x\,\varphi\, S_2[\Psi]\, \varphi \right),
\end{equation}
yields the standard result
\begin{equation}
    \Gamma^{(1)}[\Psi]
    = \frac{i}{2}\, \ln \det S_{2}[\Psi].
\end{equation}
To extract physical information from the effective action, such as the vacuum structure, we consider the effective potential, $V_{\text{eff}}$, defined as the effective action evaluated for constant background fields. This removes all dependence on derivatives and isolates the quantum-corrected potential energy density. By expanding the effective action around a background field and neglecting the slowly varying background derivatives, the leading term in the derivative expansion is obtained
\begin{equation}
    \Gamma[\Psi] = 
    \int d^{4}x\,
    \sqrt{g}
    \left[
    - V_{\mathrm{eff}}(\Psi)
    + \mathcal{O}(\partial \Psi)\right].
\end{equation}
Thus, the effective potential admits a loop expansion of the form $V_{\mathrm{eff}}(\Psi) = \sum_{i=0} V^{(i)}(\Psi)$, where $V^{(0)}$ is the classical contribution, and $V^{(1)}$ represents the one-loop quantum correction.

We perform the Wick rotation $t = -i\tau$ in order to obtain a convergent functional integral in Euclidean coordinates. Under this transformation, the one-loop contribution can be expressed as
\begin{equation}
    V^{(1)}(\Psi) = 
    -\frac{1}{\mathcal{V}_{E}}\ln\int\mathcal{D}\varphi\exp\left(
    -\frac{1}{2}\int\,d\tau d^3x\,\varphi\,S_2(\Psi)\,\varphi\right)
    \,,
\end{equation}
where $\mathcal{V}_{E} = \int d\tau\, d^{3}x\sqrt{g}$ is the volume and $S_{2}(\Psi)$ represents the Euclidean elliptic operator, which we will call $\hat{Q}$. In our case, as follows from Eq.~(\ref{action1}), the elliptic differential operator is given by
\begin{eqnarray}
    \hat{Q}=-\partial^{2}_{\tau} - \Delta + m^{2} + \frac{g}{2}\Psi^{2}\,.
\end{eqnarray}
Therefore, the problem reduces to computing the determinant of the operator. To evaluate this determinant, we employ the generalized  $\zeta$-function regularization method. In this framework, the determinant is given by
\begin{equation}
    \ln \det\hat{Q} =
    - \left.\zeta'_{\hat{Q}}(s) \right|_{s=0},
\end{equation}
with
\begin{equation}\label{zetaQ}
    \zeta_{\hat{Q}}(s) = \sum_{n} \omega_{n}^{-s},
\end{equation}
where the set $\{\omega_{n}\}$ are the eigenvalues of the total operator $ \hat{Q}$. Thus, the one-loop contribution in terms of the generalized $\zeta$-function can be written as
\begin{eqnarray}\label{VQ1}
    V^{(1)}(\Psi) =
    -\frac{1}{2\mathcal{V}_E}
    \left[\zeta'(0) + \zeta(0)\ln\mu^2
    \right]
    \,,
\end{eqnarray}
where the parameter $\mu$ has dimensions of mass is eliminated through the renormalization procedure.

To determine the observable values of the coupling constants and avoid divergences, it is necessary to implement the renormalization process. This procedure is carried out by calculating:
\begin{eqnarray}\label{Recond}
    \left.\frac{d^{4}V_{\text{eff}}(\Psi)}{d\Psi^{4}}\right|_{\hat{\Psi}}=g\,,\qquad 
    \left.\frac{d^{2}V_{\text{eff}}(\Psi)}{d\Psi^{2}}\right|_{\Psi_{0}}=m^{2},
\end{eqnarray}
where $\hat{\Psi}$ is a parameter with dimensions of mass, which can be zero in the case of a massive field, and $\Psi_0$ is the field configuration that minimizes the effective potential, satisfying $V'_{\text{eff}}(\Psi_0) = 0$, together with the stability condition $V''_{\text{eff}}(\Psi_0) > 0$. The latter ensures that $\Psi_0$ corresponds to a local minimum and is physically related to the positivity of the squared mass of scalar fluctuations around the vacuum. This configuration defines the ground state of the theory. A consistent renormalization of the vacuum energy is imposed by requiring $V_{\text{eff}}(\Psi_0) = 0$, thereby fixing the zero-point energy. After renormalization and identification of the vacuum state, physical observables such as the Casimir energy can be computed from the quantum corrections encoded in the renormalized one-loop effective potential.

Within this framework, we compute the one-loop quantum corrections to the effective potential. After implementing the appropriate renormalization procedure, we determine the Casimir energy between two neutral, parallel conducting slowly varying rough plates, satisfying Dirichlet boundary conditions and with finite temperature. In the following subsection, we analyze how to model this roughness and its impact on the generalized $\zeta$-function, which plays a central role in our approach.


\subsection{Geometric configuration and eigenvalues problem}

The geometry of the parallel conducting plates is described in Cartesian coordinates, where the $w$ coordinate is bounded by $0 \leq w \leq a + f(x,y)$, with $(x,y) \in \mathbb{R}^2$. The parameter $a$ denotes the separation between the plates, while the function $f(x,y)$ encodes the boundary roughness, assumed to satisfy $f(x,y) \ll a$ \cite{Droguett:2025frq,Borquez:2023ajx}.
To make the effect of roughness explicit as a perturbation of the geometry, we perform a change of variable in the $w$ coordinate such that the boundaries become flat, we define
\[
w = \sigma \left(1 + \frac{f(x,y)}{a}\right), \qquad 0 \leq \sigma \leq a.
\]
In the perturbative regime, roughness contributions from both plates can be treated as linear perturbations, allowing for additive or canceling effects depending on their functional form. With this choice of coordinates, the contribution of the roughness is explicitly retained. The spatial metric associated with these new coordinates is
\begin{eqnarray}
    g_{ij} &=& 
    \left(
    \begin{array}{ccc}
    1 + 
    \left(
    \sigma/a
    \right)^2(\partial_x f)^2
    & \left(
    \sigma/a
    \right)^2 \partial_x f \partial_yf 
    & \frac{\sigma}{a} \left(1+\frac{f}{a}\right) \partial_xf
    \\
    \left(
    \sigma/a
    \right)^2 \partial_x f \partial_y f  
    & 1 + 
    \left(
    \sigma/a
    \right)^2
    (\partial_yf)^2 
    & \frac{\sigma}{a} 
    \left(
    1+\frac{f}{a}
    \right) \partial_yf
    \\
    \frac{\sigma}{a}
    \left(
    1+\frac{f}{a}
    \right) \partial_xf
    & \frac{\sigma}{a}  
    \left(
    1+\frac{f}{a}
    \right) \partial_yf
    & \left(
    1+\frac{f}{a}
    \right)^2 
    \\
    \end{array}
    \right)
    \,.
    \label{metric}
\end{eqnarray}
The Laplace-Beltrami operator on a scalar field can be constructed and takes the general form
\begin{eqnarray}\label{operatorLB}
    \Delta\Phi
    &=&
    \Delta_x\Phi
    + \Delta_y\Phi
    + \frac{1}{(a+f)^2}\left[
    \sigma^{2}\left((\partial_{x}f)^{2} + (\partial_{y}f)^{2}\right)
    + a^2\right]\Delta_{\sigma}\Phi
    \nonumber
    \\&&
    + \frac{\sigma}{(a+f)^2}\left[
    2\left((\partial_{x}f)^{2} + (\partial_{y}f)^{2}\right)
    - (a+f)\left(\Delta_{x}f + \Delta_{y}f\right)
    \right]\partial_{\sigma}\Phi
    \nonumber
    \\&&
    - \frac{2\sigma}{a+f}\left(
    \partial_{x}f\Delta_{x\sigma}\Phi
     + \partial_{y}f\Delta_{y\sigma}\Phi\right)
     \,.
\end{eqnarray}
To simplify the analysis, we perform a second change of variables to introduce dimensionless coordinates:
\begin{equation}
    \begin{split}
    x & = u_1 L_1\,, \qquad -1/2\leq u_1\leq 1/2\,,
    \\
    y & = u_2 L_2\,, \qquad  -1/2\leq u_2\leq 1/2\,,
    \\
    \sigma & = va \,,\qquad  \qquad 0\leq v\leq 1\,.
    \end{split}
    \label{finalchangeofvariable}
\end{equation}
As mentioned earlier, the function $f$ (from now on $\hat{f}=f(u_iL_i)$, with $i=1,2$ in the new coordinates) has a perturbative nature, therefore, it allows us to arrive at a useful simplification for the operator (\ref{operatorLB}) by assuming that the roughness profile varies smoothly along the plates over the characteristic length scales $L_{i}$, which therefore play the role of effective correlation lengths for the surface deformation. The validity of the approximation is controlled by the adiabatic conditions $a/L_i\ll 1$, $|\hat f|/a\ll 1$, $|\partial \hat f|\ll 1$. Under these assumptions, derivative contributions such as $\partial_i \hat f$ and $\partial_i^2 \hat f$ are suppressed with respect to the leading geometric corrections associated with the local deformation amplitude (Because both parameters $L_{1}$ and $L_{2}$ are infinitely large compared to the distance $a$ and the function $\hat{f}$, we set $L_{1}=L_{2}=L$). For the remaining terms, in a compact form, we have
\begin{eqnarray}
    \Delta\Phi
    =
    \left(\frac{1}{L^2}\Delta_{u_{i}}
    + \frac{1}{a^2}\mathcal{N}(u_{i})\,\Delta_v
    \right)\Phi\,,
\label{Delta}
\end{eqnarray}
where, by considering a perturbative expansion of $\hat{f}$ up to any order, $\mathcal{N}(u_{i})$ is defined by
\begin{eqnarray}
    \mathcal{N}(u_{i}) =
    1
    - 2\frac{\hat{f}(u_{i})}{a}
    + 3\left(\frac{\hat{f}(u_{i})}{a}\right)^{2}
    - 4\left(\frac{\hat{f}(u_{i})}{a}\right)^{3}
    + \ldots
    \label{N}
\end{eqnarray}
with $\ldots$ denoting higher-order contributions in the perturbative expansion of the roughness function $\hat{f}(u_i)$. With this change of variables, the spatial geometry of the plates is fully specified. The Euclidean volume in (\ref{VQ1}) is given by $\mathcal{V}_E = \xi \, \mathcal{V}_3$, where $\mathcal{V}_3 = \int d^3x \, \sqrt{g}$. Therefore\footnote{Note that the spatial volume depends on the roughness defined in the geometry of the plates.},
\begin{eqnarray}\label{Vol2}
    \mathcal{V}_{3} =
    L^{2}\left(a 
    + \int_{-1/2}^{1/2}\int_{-1/2}^{1/2}\hat{f}(u_{i})\,du_{i}\right).
\end{eqnarray}

Returning to the general eigenvalue problem, the full spectrum associated with the operator $\hat{Q}$ is given by
\begin{eqnarray}
    \left(-\partial_{\tau}^{2} + \hat{\mathcal{P}}\right)\Phi 
    =
    \omega\,\Phi\,,
\end{eqnarray}
where $\hat{\mathcal{P}} = -\Delta + m^{2} + \frac{g}{2}\Psi^{2}$, and $\omega$ denotes its eigenvalue. The eigenvalues associated with the temporal sector arise from imposing periodic boundary conditions along the Euclidean time direction. In the limit $\xi \to \infty$, one recovers the zero-temperature case. The boundary conditions imposed on the spatial sector are of Dirichlet type,
\begin{eqnarray}
    \Phi(\tau,u_{i},0) 
    =
    \Phi(\tau,u_{i},1)
    = 0\,.
\label{DBC}
\end{eqnarray}
The full eigenvalue is
\begin{eqnarray}\label{EingTotal}
    \omega_{l,k_{i},n} =
    \left(\frac{2\pi l}{\xi}\right)^{2}
    + \lambda_{k_{i},n}^{2}\,,
\end{eqnarray}
with $l \in \mathbb{Z}$, and $\lambda_{k_{i},n}$ denoting the eigenvalues associated with the spatial operator $\hat{\mathcal{P}}$.

Since (\ref{EingTotal}), we use the integral form of the $\zeta$-function (see Eq. (\ref{zetaQ})) to rewrite the spectral function as
\begin{eqnarray}
    \zeta_{\hat{Q}}(s) =
    \frac{1}{\Gamma(s)}\int_{0}^{\infty}dt\,t^{s-1}
    \sum_{l=-\infty}^{\infty}
    \sum_{n,k_{i}=1}^{\infty}
    \exp\left\{
    -t\left[\left(\frac{2\pi l}{\xi}\right)^{2}
    + \lambda_{k_{i},n}^{2}\right]
    \right\}.
\end{eqnarray}
Then, a suitable representation of the $\zeta$-function is obtained by applying the Poisson resummation formula
\begin{eqnarray}\label{ZeFirst}
    \zeta_{\hat{Q}}(s) &=&
    \frac{\xi}{\sqrt{4\pi}}\frac{\Gamma(s-1/2)}{\Gamma(s)}\,\zeta_{\hat{\mathcal{P}}}(s-1/2)
    \nonumber\\
    &&
    + \frac{\xi}{\sqrt{\pi}\,\Gamma(s)}
    \sum_{l=1}^{\infty}
    \sum_{n,k_{i}=1}^{\infty}
    \int_{0}^{\infty}dt\,t^{s-3/2}
    \exp\left[
    - \frac{\xi^{2}l^{2}}{4t}
    - \lambda_{k_{i},n}^{2}t \right].
\end{eqnarray}
This representation allows us to decouple the spatial sector from the temperature-dependent terms. The first term ($l=0$) corresponds to a purely spatial contribution, where a Riemann $\zeta$-function is defined and potential divergences may arise. The second term contains all temperature-dependent corrections. This decomposition is particularly useful for analytically studying both the low- and high-temperature regimes, allowing the extraction of relevant thermodynamic quantities. In the present work, we focus on the low-temperature limit.


\subsection{Contour integration method}
\label{Sec2.2}

In general, the eigenvalue problem for the spatial operator does not admit a closed-form spectral condition due to the presence of boundary roughness, which significantly complicates the structure of the differential operator $\hat{\mathcal{P}}$. For this reason, it is convenient to employ the contour integration method. In this approach, the eigenvalues are determined by the zeros of an associated spectral function, which can be written as \cite{Kirsten:2007ev,Jeffres:2012de}
\begin{eqnarray}\label{F}
    \mathcal{F}_{j}(\rho)=\frac{\phi_{j}(\rho;1)}{\phi_{j}(0;1)},
\end{eqnarray}
where $\rho$ corresponds to the main eigenvalue and the subindex $j$ represents any other type of eigenvalue present associated with the other coordinates. By the residue theorem, the $\zeta$-function can be expressed as a contour integral,
\begin{eqnarray}
    \zeta(s) =
    \frac{1}{2\pi i}\sum_{j}
    \int_{\mathcal{M}}\rho^{-2s}\frac{d}{d\rho}\ln \mathcal{F}_{j}(\rho)\,d\rho,
\end{eqnarray}
where $\mathcal{M}$ is a contour enclosing all nonvanishing eigenvalues $\rho \neq 0$. The aim of this section is to determine the explicit form of the spectral function $\mathcal{F}_{j}(\rho)$ adapted to our problem.

The spatial sector of the general solution $\Phi(\tau,u_i,v)$, which we denote by $\phi(u_i,v)$, is obtained using the WKB method. To simplify the analysis, we assume that the geometric perturbation $\mathcal{N}$ depends only on the coordinate $u_{1}$. The corresponding eigenvalue problem is then given by
\begin{eqnarray}
    \hat{\mathcal{P}}\,\phi(u_{i},v)
    = \left(-\Delta 
    + m^{2} 
    + \frac{g}{2}\Psi^{2}\right)
    \phi(u_{i},v) = 
    \lambda_{k,n}^{2}\,\phi(u_{i},v)\,,
\end{eqnarray}
where $\lambda_{k,n}$ represents the spatial eigenvalue (with $k,n$ denoting the eigenvalues of the free coordinates)\footnote{By choosing that $\mathcal{N}$ is only dependent on $u_{1}$, the coordinates $u_{2}$ and $v$ are not affected by the roughness perturbations, generating free eigenvalues in the spectral function, $k$ and $n$, respectively.}; the Laplace Beltrami operator is given in (\ref{Delta}) considering $\mathcal{N}$ only depends of $u_{1}$, and we define $M^{2} = m^{2} + \frac{g}{2}\Psi^{2}$.  The corresponding ansatz for the solution is
\begin{eqnarray}
    \phi(u_{i},v) =
    \chi(u_{1})
    \sin(n\pi v)
    e^{iku_{2}}
    \,,
    \label{Sol1}
\end{eqnarray}
with $n \in \mathbb{N}$, which leads to the following differential equation
\begin{eqnarray}
    \chi''(u_{1})
    + \left[
    (\lambda^{2}
    - k^{2}
    - M^{2})L^{2}
    - \left(\frac{n\pi L}{a}\right)^{2}
    \mathcal{N}(u_{1})
    \right]\chi(u_{1})= 0,
\end{eqnarray}
where $k = \frac{\pi q}{L}$, with $q \in \mathbb{Z}$. We rewrite $\rho^{2} = \lambda^{2} - k^{2}$, where $\rho$ corresponds to the eigenvalue associated with the coordinate $u_{1}$. As part of the procedure to determine (\ref{F}), we perform an analytic continuation to the complex plane, $\rho^{2} \to -\pi^{2} n^{2} z^{2}$ \cite{Jeffres:2012de}. Then,
\begin{eqnarray}
    \chi''(u_{1})
    - L^{2}\left[
    \pi^{2}n^{2}\left(z^{2} +\frac{\mathcal{N}(u_{1})}{a^{2}}\right)
    + M^{2}
    \right]\chi(u_{1})=0
    \,.
\end{eqnarray}
The proposed solution is obtained using the WKB method and takes the general form\footnote{ At this point, we have shifted the edges of the solution to $u_{1}\in \{0,1\}$, keeping the magnitude of the separation distance unchanged. For our problem, this change has no implications; however, it must be taken into account once the shape of the roughness function is specified.}
\begin{eqnarray}\label{Rsol}
    \chi(u_{1}) =
    \exp
    \left[\int_{0}^{u_{1}}R(z,n;\beta)\,d\beta\right].
\end{eqnarray}
The function $R(z,n;u_{1})$, once the integral is evaluated, admits an asymptotic expansion in $n$ of the form
\begin{eqnarray}
    R_{\pm}(z,n;u_{1}) =
    \pm r_{-1}(z;u_{1})\,n 
    + r_{0}(z;u_{1}) 
    \pm \frac{r_{1}(z;u_{1})}{n} + \ldots
\end{eqnarray}
By collecting terms order by order in $n$, we obtain the following solutions:\footnote{The treatment to determine the solutions of the asymptotic expansion is carried out recursively, as was done in \cite{Jeffres:2012de}. Our analysis is analogous to that done by the authors, where the surface topology is determined by a general two-dimensional metric.}
\begin{eqnarray}
    r_{-1}(z;u_{1}) &=& 
    \pi L\left(z^{2} 
    + \frac{\mathcal{N}(u_{1})}{a^{2}}\right)^{1/2},
    \nonumber\\
    r_{0}(z;u_{1}) &=& 
    - \frac{1}{2\,r_{-1}(z;u_{1})}\frac{\partial} {\partial u_{1}}r_{-1}(z;u_{1}),
    \\
    r_{1}(z;u_{1}) &=&
    \frac{1}{2r_{-1}(z;u_{1})}
    \left(M^{2}L^{2}-r^{2}_{0}(z;u_{1})
    - \frac{\partial}{\partial u_{1}}r_{0}(z;u_{1})\right).
    \nonumber
\end{eqnarray}
The solution $r_{0}(z;u_{1})$ consists of terms of the form $\partial \hat{f}, \partial^{2}\hat{f}$. As mentioned above, in the adiabatic regime derivative contributions are suppressed compared to the leading amplitude corrections. Therefore, at leading order, we neglect gradient contributions, that is, $r_{0}(z;u_{1})=0$. Then, the solution is
\begin{eqnarray}
    \chi(u_{1}) =
    A\exp\left[\int_{0}^{u_{1}}R_{+}(z,n;\beta)\,d\beta\right]
    + B\exp\left[\int_{0}^{u_{1}}R_{-}(z,n;\beta)\,d\beta\right]
    \,,
\end{eqnarray}
with $A$ and $B$ constants. Imposing the boundary conditions given in (\ref{DBC}), $\chi(0)=\chi(1)=0$, we obtain an eigenvalue function of the form
\begin{eqnarray}
    \mathcal{F}_{n}(z) =
    e^{\int_{0}^{1}R_{1}(z,n;\beta)\,d\beta}
    \left(
    1 - e^{-2\int_{0}^{1}R_{1}(z,n;\beta)\,d\beta}\right)
    \,,
\end{eqnarray}
where 
\begin{eqnarray}\label{R1}
    R_{1} (z,n;\beta) =
    n\pi L\left(z^{2} + \frac{\mathcal{N}(\beta)}{a^{2}}\right)^{1/2}
    + \frac{M^{2}L}{2n\pi}\left(z^{2} + \frac{\mathcal{N}(\beta)}{a^{2}}\right)^{-1/2}.
\end{eqnarray}
As defined previously, the total eigenvalue in the spatial sector is $\lambda^{2} = \rho^{2} + k^{2}$. The rotation applied to $\rho$ suggests introducing a similar transformation for $k$, namely $k = \pi n \eta z$, where $0 \leq z < \infty$ and $0 \leq \eta \leq 1$. Hence, we have\footnote{Here it is convenient to move to the continuous limit of the eigenvalue $k$, that is $\sum _k\rightarrow \left(\frac{L}{2\pi}\right)\int dk$.} 
\begin{eqnarray}
    \zeta_{\hat{\mathcal{P}}}(s) &=&
    \frac{\sin\pi s}{\pi}\left(\frac{L}{\pi}\right)
    \sum_{n=1}^{\infty}(\pi n)^{-2s+1}
    \int_{0}^{1}(1-\eta^{2})^{-s}\,d\eta
    \nonumber\\
    &&
    \times
    \int_{0}^{\infty}
    z^{-2s+1}\left(\frac{\partial}{\partial z} - \frac{\eta}{z}\frac{\partial}{\partial\eta}\right)\ln\mathcal{F}_{n}(z)\,dz.
\end{eqnarray}
At this point, several aspects must be taken into account. First, from (\ref{F}) and the solution (\ref{Sol1}), we observe that the function $\mathcal{F}_{n}$ is effectively determined by the solution at the coordinate $u_{1} \in \{0,1\}$. Second, the function $\mathcal{F}_{n}$ does not depend explicitly on $\eta$, therefore, its derivative with respect to $\eta$ vanishes. Finally, taking the logarithm of $\mathcal{F}_{n}$, we obtain
\begin{eqnarray}\label{Alog}
    \int_{0}^{1}R_{1}(z,n;\beta)\,d\beta
    + \ln\left[1 - e^{-2\int_{0}^{1}R_{1}(z,n;\beta)\,d\beta}\right]
    - \ln\chi_{n}(0;1)\,.
\end{eqnarray}
We observe that the last term does not depend on $z$, while the second term contains the solution $R_{1}$, which is proportional to $L$ (see Eq.~(\ref{R1})). As part of the renormalization procedure, we take the limit $L \to \infty$, consequently, the first term in (\ref{Alog}) is the only one that contributes to the result of interest. Hence,
\begin{eqnarray}\label{zeta0}
    \zeta_{\hat{\mathcal{P}}}(s) &=&
    \frac{\sin\pi s}{\pi}\left(\frac{L}{\pi}\right)
    \sum_{n=1}^{\infty}(\pi n)^{-2s+1}
    \int_{0}^{1}(1-\eta^{2})^{-s}\,d\eta
    \nonumber\\
    &&
    \times\int_{0}^{\infty}
    z^{-2s+1}\frac{\partial}{\partial z}\int_{0}^{1}R_{1}(z,n;\beta)\,d\beta\,dz.
\end{eqnarray}
Using the following standard formulas in (\ref{zeta0})
\begin{eqnarray}
    \frac{\sin\pi s}{\pi}=\frac{1}{\Gamma(s)\Gamma(1-s)}\,,
\end{eqnarray}
\begin{eqnarray}
    \int_{0}^{1}(1-\eta^{2})^{-s}\,d\eta =\frac{\sqrt{\pi}}{2}\frac{\Gamma(1-s)}{\Gamma\left(\frac{3}{2}-s\right)}\,,
\end{eqnarray}
\begin{eqnarray}
\int_{0}^{\infty}\frac{z^{-s}}{(z+1)^{p}}dz=\frac{\Gamma(1-s)\Gamma(s+p-1)}{\Gamma(p)}\,,
\end{eqnarray}
we obtain
\begin{eqnarray}\label{zeta1}
    \zeta_{\hat{\mathcal{P}}}(s) &=&
    \frac{L^{2}}{4\pi^{2s-1}}\frac{\Gamma(s-1)}{\Gamma(s)}\,\zeta_{R}(2s-2)\,a^{2s-2}
    \int_{0}^{1}\mathcal{N}^{1-s}(\beta)\,d\beta
    \nonumber\\
    &&
    - \frac{M^{2}L^{2}}{4\,\pi^{2s+1}}\,
    \zeta_{R}(2s)\,a^{2s}
    \int_{0}^{1}\mathcal{N}^{-s}(\beta)\,d\beta
    \,.
\end{eqnarray}
In Eq.~(\ref{ZeFirst}), we perform the shift $s \to s - \tfrac{1}{2}$ as required in (\ref{zeta1}). The one-loop potential (\ref{VQ1}) involves the derivative of $\zeta_{\hat{Q}}(s)$ and must be evaluated at $s = 0$. This evaluation leads to divergences, therefore, we perform a series expansion in $s$. For the second term in (\ref{VQ1}), after expanding in $s$, we find that $\zeta_{\hat{Q}}(0) = 0$. From these results, the one-loop correction to the effective potential reads
\begin{eqnarray}\label{V1Q2}
    V_{\hat{Q}}^{(1)} &=&
    - \frac{1}{2\mathcal{V}_{3}}
    \left[\frac{\pi^{2}L^{2}}{720\,a^{3}}
    \int_{0}^{1}\mathcal{N}^{3/2}(\beta)\,d\beta
    -
    \frac{M^{2}L^{2}}{48\,a}\int_{0}^{1}\mathcal{N}^{1/2}(\beta)\,d\beta
    \right.
    \nonumber\\
    &&
    \left.
    +
    \frac{1}{\sqrt{\pi}}\sum_{l,n,k_{i}=1}^{\infty}\int_{0}^{\infty}dt\,t^{-3/2}\exp\left(-\frac{\xi^{2}l^{2}}{4t}-\lambda_{k_{i},n}^{2}t \right) 
    \right]\,.
\end{eqnarray}


\subsection{Low-temperature limit}

The zero-temperature contribution and the thermal correction probe different regions of the spectrum. The contour integral representation of the vacuum contribution requires an asymptotic description of the full spectrum, for which the WKB approximation is appropriate. In contrast, in the low-temperature limit $\xi\gg a$, the thermal contribution is exponentially dominated by the lowest eigenvalues. Therefore, a first-order perturbative treatment of the roughness is sufficient to determine the leading thermal correction.
The last term in (\ref{V1Q2}) contains the entire temperature dependence. One of our goals is to analytically develop the temperature contribution, and for this purpose, we restrict ourselves to analyzing the low-temperature case \cite{Bordag:2001qi}. 
The integral over $t$ in (\ref{V1Q2}) can be performed directly and we can solve the sum over $l$. We obtain
\begin{eqnarray}\label{T02}
    - \frac{2\sqrt{\pi}}{\xi}\sum_{n,k_{i}=1}^{\infty}\ln\left( 1 - e^{-\lambda_{k_{i},n}\xi}\right).
\end{eqnarray}
As previously established, the eigenvalue $\lambda_{k_{i},n}$ is associated with the spatial operator $\hat{\mathcal{P}}$. Thus far, we have not specified the structure of this eigenvalue. Since the low-temperature contribution is dominated by the lowest eigenvalues of the spatial operator, it is sufficient to employ the perturbative approximation developed in \cite{Droguett:2025frq}. To find an expression for $\lambda_{k_{i},n}$ in the presence of rough plates, we use perturbation theory. Within this framework, the spectrum is obtained at first order by treating the roughness as a small correction that modifies the free spatial operator. Therefore, the general structure of the eigenvalue is $\lambda=\lambda^{(0)} + \lambda^{(1)}$, where $\lambda^{(0)}$ represents the free spectrum, and $\lambda^{(1)}$ defines the spectral contribution of smooth roughness as perturbative corrections. According to the method used in \cite{Droguett:2025frq}, the total spatial eigenvalue is
\begin{eqnarray}\label{FirstOrder}
    \lambda_{k_{1},k_{2},n}^{2} = 
    \left(\frac{\pi k_{1}}{L}\right)^{2} 
    + \left(\frac{\pi k_{2}}{L}\right)^{2}
    +\left(\frac{n\pi}{a}\right)^{2}\int_{0}^{1}\mathcal{N}(\beta)\,d\beta
    + M^{2},
\end{eqnarray}
where $k_{1}$ and $k_{2}$ are the eigenvalues corresponding to the original coordinate pair $(x,y)$ (Again, for the eigenvalues $k_{1},k_{2}$ we work in the continuous limit). Equation (\ref{FirstOrder}) should be understood as the first-order perturbative approximation to the same spatial spectrum analyzed through the WKB method in Sect. \ref{Sec2.2}, considering now that we are working in the regime $\xi M\gg1$ and $\xi\gg a$.

The current calculation is valid in the principal order in the adiabatic roughness expansion, first order in spectrum perturbations and principal order in the low temperature expansion $\xi\gg a$.
In this limit, the dominant contribution to the exponential in Eq. \eqref{T02} arises from the infrared regime, characterized by small values of $k_1, k_2$ and low-frequency modes $n$. Given that the eigenvalue $\lambda_{k_1, k_2, n}$ appears under a square root in Eq. \eqref{T02}, we perform a Taylor expansion around $k_1, k_2 \approx 0$. Furthermore, the convergent asymptotic behavior of the exponential justifies the approximation $\ln(1+x) \approx x$ for $x \ll 1$. Consequently, substituting these approximations into Eq. \eqref{T02}, we obtain
\begin{eqnarray}
    \frac{2\sqrt{\pi}}{\xi}\left(\frac{L}{2\pi}\right)^{2}\int_{-\infty}^{\infty}\int_{-\infty}^{\infty}dk_{1}dk_{2}
    \sum_{n=1}^{\infty}
    \exp\left[
    -\xi\left(\mathcal{R}_{n}^{1/2}
    +
    \frac{k_{1}^{2}+ k_{2}^{2}}{2
    \mathcal{R}_{n}^{1/2}}\right)
    \right],
\end{eqnarray}
with 
\begin{eqnarray}
    \mathcal{R}_{n}=\left(\frac{n\pi}{a}\right)^{2}\int_{0}^{1}\mathcal{N}(\beta)\,d\beta
    + M^{2}.
\end{eqnarray}
The integration over the momenta $k_1$ and $k_2$ can be performed analytically. Furthermore, since $n \neq 0$, the sum over $n$ is dominated by the $n = 1$ term, which constitutes the leading-order contribution. Finally, substituting these results into Eq. (\ref{V1Q2}), the one-loop effective potential takes the form
\begin{eqnarray}\label{V1Q3}
    V_{\hat{Q}}^{(1)} &=&
    - \frac{L^{2}}{2\mathcal{V}_{3}}
    \left[\frac{\pi^{2}}{720\,a^{3}}
    \int_{0}^{1}\mathcal{N}^{3/2}(\beta)\,d\beta
    -
    \frac{M^{2}}{48\,a}\int_{0}^{1}\mathcal{N}^{1/2}(\beta)\,d\beta
    +
    \frac{\mathcal{R}_{1}^{1/2}}{\pi\xi^{2}}
    \exp\left(-\xi\,
    \mathcal{R}_{1}^{1/2}\right)
    \right].
    \nonumber\\
\end{eqnarray}
In the temperature sector, this equation shows an expression that can be suppressed in the limit $\xi\rightarrow\infty$, however, we have arrived at an analytical expression where its asymptotic behavior at low temperature is determined by small values of the eigenvalue spectrum of the spatial sector. In what follows, we shall apply the aforementioned renormalization procedure to evaluate the one-loop renormalized effective potential.


\section{Renormalization process}

\subsection{Counterterms and vacuum stability}
\label{sec3.1}
In Eq. \eqref{V1Q3}, we established the expression for the unrenormalized one-loop effective potential. By incorporating the physical considerations and limits discussed in the previous sections, we arrive at the following result
\begin{eqnarray}
    V_{\text{eff}}(\Psi) &=&
    \frac{m^{2} + \delta_{m}}{2}\Psi^{2} 
    + \frac{g + \delta g}{4!}\Psi^{4} 
    + \delta c
    - \frac{L^{2}}{2\mathcal{V}_{3}}
    \left[\frac{\pi^{2}}{720\,a^{3}}
    \int_{0}^{1}\mathcal{N}^{3/2}(\beta)\,d\beta
    \right.
    \nonumber\\
    &&
    \left.
    -
    \frac{M^{2}}{48\,a}\int_{0}^{1}\mathcal{N}^{1/2}(\beta)\,d\beta
    +
    \frac{\mathcal{R}_{1}^{1/2}}{\pi\xi^{2}}
    \exp\left(-\xi\,
    \mathcal{R}_{1}^{1/2}\right)
    \right].
\end{eqnarray}
The renormalization process is implemented through the set of conditions defined in Eqs. \eqref{Recond}. It is important to note that the vacuum state $\Psi_{0}$ is determined by the stationarity condition $V^{'}_{\text{eff}}(\Psi_{0})=0$, while the stability of this state is guaranteed by the condition $V^{''}_{\text{eff}}(\Psi_{0})>0$. For the present analysis, we assume that $\Psi_{0}=0$ corresponds to the stable vacuum of the theory. To determine the counterterm $\delta c$, we impose the condition $V_{\text{eff}}(\Psi_0)=0$. In the limit $a \to \infty$, we find that $\delta c = 0$, a result stemming from the fact that the potential is scaled by the volume $\mathcal{V}_{3}$, which depends directly on the parameter $a$ as shown in Eq. \eqref{Vol2}. A similar procedure applied to the remaining counterterms yields consistent results: by enforcing the first condition in \eqref{Recond} and taking the large $a$ limit, the coupling constant counterterm vanishes, $\delta g = 0$, while the second condition implies $\delta m = 0$. Consequently, the renormalized effective potential reduces to the following form

\begin{eqnarray}\label{VR}
    V^{R}_{\text{eff}}(\Psi) 
    &=&
    \frac{m^{2}}{2}\Psi^{2} 
    + \frac{g}{4!}\Psi^{4} 
    - \frac{1}{2\left(a+\int_{0}^{1}\hat{f}(\beta)d\beta\right)}
    \left[\frac{\pi^{2}}{720\,a^{3}}
    \int_{0}^{1}\mathcal{N}^{3/2}(\beta)\,d\beta
    \right.
    \nonumber\\
    &&
    \left.
    -
    \frac{M^{2}}{48\,a}\int_{0}^{1}\mathcal{N}^{1/2}(\beta)\,d\beta
    +
    \frac{\mathcal{R}_{1}^{1/2}}{\pi\xi^{2}}
    \exp\left(-\xi\,
    \mathcal{R}_{1}^{1/2}\right)
    \right].
\end{eqnarray}
This result demonstrates that the renormalized effective potential depends explicitly on both the temperature and the smooth geometric roughness defined in Eq. \eqref{N}. In particular, no remnant terms depending on $\mathcal{N}$ or $\xi$ appear in the counterterms. We emphasize that no renormalization conditions were imposed on these parameters, nor was it necessary to cancel these variables to complete the renormalization procedure. At the minimum $\Psi_{0}$, the potential retains terms originating from the topology; however, in the limit $\xi \to \infty$, the last term vanishes, indicating that the potential depends solely on the plate separation and the surface roughness.
Furthermore, in the limit $a \to \infty$, all terms vanish, which is consistent with the fact that the vacuum energy must be zero for infinitely separated plates, thus recovering the results of a free Minkowski spacetime. Another significant finding is that no explicit counterterms were required, suggesting that this particular approach is finite. This conclusion is specific to the problem addressed here. The WKB approximation allows obtaining an analytical representation of the eigenvalue spectrum, which typically encodes the ultraviolet structure of the theory. Therefore, the finiteness of the effective potential is achieved by constructing the spectral function $\zeta$ together with the renormalization conditions adopted in this work. Specifically, if any term had survived the limit $a \to \infty$ during the renormalization process, counterterms would have been necessary. However, the structure of the spectral solution prevents any simplification of the parameter $a$ that would lead to topology-independent terms. Notably, these conclusions remain valid without specifying the explicit form of the function $\hat{f}$.

To further explore the vacuum structure, we consider the massless limit $m=0$. This choice corresponds to the asymptotic regime of the theory, where the characteristic energy scales of the system dictated by the temperature  and the plate separation are significantly larger than the rest mass of the field. In this limit, the effective potential is primarily governed by radiative corrections and geometric constraints, allowing us to isolate the subtle effects of the topological roughness. Under this assumption, the renormalized effective potential in Eq. \eqref{VR} reduces to
\begin{eqnarray}
    V^{R}_{\text{eff}}(\Psi) &=& 
    \frac{g}{4!}\Psi^{4} 
    - \frac{1}{2\left(a+\int_{0}^{1}\hat{f}(\beta)d\beta\right)}
    \left[\frac{\pi^{2}}{720\,a^{3}}
    \int_{0}^{1}\mathcal{N}^{3/2}(\beta)\,d\beta
    -
    \frac{g\Psi^{2}}{96\,a}\int_{0}^{1}\mathcal{N}^{1/2}(\beta)\,d\beta
    \right.
    \nonumber\\
    &&
    \left.
    +
    \frac{1}{\pi\xi^{2}}
    \left(
    \frac{\pi^{2}}{a^{2}}\int_{0}^{1}\mathcal{N}(\beta)\,d\beta
    + \frac{g\Psi^{2}}{2}\right)^{1/2}
    \exp\left(-\xi\left(
    \frac{\pi^{2}}{a^{2}}\int_{0}^{1}\mathcal{N}(\beta)\,d\beta
    + \frac{g\Psi^{2}}{2}\right)^{1/2}\right)
    \right].
    \nonumber\\
\end{eqnarray}
To determine the vacuum expectation value of the field, we adopt a perturbative expansion in the coupling constant $g$, assuming the weak coupling regime ($g \ll 1$). Then, by imposing the stationary condition, we find
\begin{eqnarray}\label{Minimo}
    0 &=&
    \Psi \left[
    \frac{1}{6}\Psi^{2}
    + \frac{\int_{0}^{1}\mathcal{N}^{1/2}(\beta)d\beta}{96a\left(a+\int_{0}^{1}\hat{f}(\beta)d\beta\right)} 
    \right.
    \nonumber\\
    &&
    \left.
   + \frac{\pi\xi 
   - a\left(
   \int_{0}^{1}\mathcal{N}(\beta)\,d\beta\right)^{-1/2}}{4\pi^{2}\xi^{2}\left(a+\int_{0}^{1}\hat{f}(\beta)d\beta
   \right)}  
   \exp\left(
    -\frac{\pi\xi}{a}\left(\int_{0}^{1}\mathcal{N}(\beta)\,d\beta\right)^{1/2}
    \right)\right]
    \,.
\end{eqnarray}
From this expression, the trivial solution is given by $\Psi_{0}=0$. The stability condition identifies the trivial solution as a valid vacuum state. This occurs because the second derivative of the effective potential at $\Psi=0$ is strictly positive
\begin{eqnarray}
    V^{''R}_{\text{eff}}(\Psi=0) &=&
    \frac{g}{\left(a+\int_{0}^{1}\hat{f}(\beta)d\beta\right)} \left[
    \frac{1}{96a}\int_{0}^{1}\mathcal{N}^{1/2}(\beta)d\beta \right.
    \nonumber\\
    &&
    \left.
    + \frac{\pi\xi - a\left(
    \int_{0}^{1}\mathcal{N}(\beta)\,d\beta\right)^{-1/2}}{4\pi^{2}\xi^{2}}  
    \exp\left(
    -\frac{\pi\xi}{a}\left(\int_{0}^{1}\mathcal{N}(\beta)\,d\beta\right)^{1/2}
    \right)\right]>0
    \,.
    \nonumber\\
\end{eqnarray}
The geometric term dominates over the temperature-dependent contributions, which decay rapidly in the low-temperature regime. Furthermore, a positive coupling constant $g$ is required to ensure the potential remains bounded from below, guaranteeing global stability.

The non-trivial solutions are obtained by solving for the terms within the brackets (\ref{Minimo}):
\begin{eqnarray}
    \Psi_{\pm} &=& 
    \pm \left\{ 
    - \frac{6}{a+\int_{0}^{1}\hat{f}(\beta)d\beta} 
    \left[
     \frac{1}{96a} \int_{0}^{1}\mathcal{N}^{1/2}(\beta)d\beta 
    \right.\right.
    \nonumber
    \\
    &&
   \left.\left.
   +  \frac{\pi\xi - a\left(
   \int_{0}^{1}\mathcal{N}(\beta)\,d\beta\right)^{-1/2}}{4\pi^{2}\xi^{2}} \exp\left(
    -\frac{\pi\xi}{a}\left(\int_{0}^{1}\mathcal{N}(\beta)d\beta\right)^{1/2}
    \right)\right]\right\}^{1/2}.
\end{eqnarray}
These solutions are physically consistent only when the argument of the square root is non-negative. However, in the low-temperature limit considered in this work, the term inside the square root remains strictly negative. Consequently, these non-trivial solutions are non-physical and do not represent a stable background for the theory. In this regime, the competition between temperature and plate geometry does not favor a phase transition, leaving the trivial solution $\Psi_0=0$ as the only stable vacuum state.


\subsection{Casimir effect and topological mass}
The vacuum energy of the system can be extracted directly from the renormalized effective potential. By evaluating the potential at the stable trivial solution $\Psi=0$, which corresponds to the symmetric phase of the theory, we define the Casimir energy density as
\begin{equation}
    \mathcal{E}_{C} = \left( a + \int_{0}^{1} \hat{f}(\beta) \, d\beta \right) \left. V^{R}_{\text{eff}}(\Psi) \right|_{\Psi=0}.
    \label{CasimirEnergyDef}
\end{equation}
This expression represents the finite vacuum response of the quantum field under the constraints of the rough boundaries. In this low-temperature regime, the Casimir energy is determined solely by the geometry and the one-loop quantum corrections, as the vacuum expectation value of the field vanishes.

Since the renormalized effective potential was specifically derived for the massless case, we set $m=0$ to maintain consistency with our previous results. Under this condition, the explicit form of the Casimir energy density is found to be
\begin{eqnarray}\label{CasEn}
    \mathcal{E}_{C} &=&
    - \frac{\pi^{2}}{1440\,a^{3}}
    \int_{0}^{1}\mathcal{N}^{3/2}(\beta)\,d\beta
    \nonumber\\
    &&
    -
    \frac{1}{2\xi^{2}a}
    \left(
    \int_{0}^{1}\mathcal{N}(\beta)\,d\beta\right)^{1/2}
    \exp\left[-\frac{\pi\xi}{a}\left(
    \int_{0}^{1}\mathcal{N}(\beta)\,d\beta\right)^{1/2}\right].
\end{eqnarray}
In the low-energy limit, the exponential term in Eq. (\ref{CasEn}) vanishes, and the Casimir energy density becomes independent of thermal effects. 
Using the perturbative expansion of $\mathcal{N}(\beta)$ introduced above (see Eq. (\ref{N})), Eq. (\ref{CasEn}) is systematically expanded in powers of the small parameter $\hat{f}/a$. This is achieved by expanding all functions of $\mathcal{N}(\beta)$, including the fractional powers and exponential factor, about the flat term configuration $\mathcal{N}=1$, while consistently retaining terms up to second order in $\hat{f}/a$. The resulting Casimir energy is\footnote{It is important to note that the first term of Eq. (\ref{CasEn}) contains the factor $\mathcal{N}^{3/2}(\beta)$ inside the integral, whereas the pre-exponential factor and the argument of the exponential both involve the square root of the integral $\int\mathcal{N}(\beta)d\beta$. When the second term of Eq. (\ref{CasEn}) is consistently expanded up to second order in $\hat{f}/a$, the combined expansions of the square-root and exponential functions must be carried out simultaneously. Since both functions depend on the same integrated quantity, the second-order expansion naturally generates cross terms that can be written as products of independent integrals. Equivalently, the quadratic correction is not obtained solely from the second-order term in the expansion of $\mathcal{N}(\beta)$, but also from the nonlinear composition of the functions appearing in Eq. (\ref{CasEn}). In particular, this procedure gives rise to the last term in (\ref{CasEnPer}).}
\begin{eqnarray}\label{CasEnPer}
    \mathcal{E}_{C} &=&
    - \frac{\pi^{2}}{1440\,a^{3}}
    -\frac{e^{-\frac{\pi\xi}{a}}}{2a\xi^{2}}
    + \left[
    \frac{\pi^{2}}{480\,a^{4}}
    + \frac{1}{a^{2}\xi^{2}}\left( 1 - \frac{\pi\xi}{a}\right)e^{-\frac{\pi\xi}{a}}
    \right]
    \int_{0}^{1}\hat{f}(\beta)d\beta
    \nonumber\\
    &&
    -
    \left[
    \frac{\pi^{2}}{240\,a^{5}}
    + \frac{1}{\xi^{2}a^{3}}\left(
    1
    - \frac{\pi\xi}{a}
    + \frac{\pi^{2}\xi^{2}}{2a^2}
    \right)e^{-\frac{\pi\xi}{a}}
    \right]\int_{0}^{1}\hat{f}^{2}(\beta)d\beta
    \nonumber\\
    &&
    + \frac{\pi}{2\xi a^{4}}e^{-\frac{\pi\xi}{a}}\left(
    \int_{0}^{1}\hat{f}(\beta)d\beta
    \right)^{2}
    + \mathcal{O}(\hat{f}^{3})
    \,.
\end{eqnarray}
The inclusion of smooth surface roughness introduces new corrections to the Casimir energy. By integrating the arbitrary function $\hat{f}$ according to Eq. (\ref{N}), we obtain an explicit expression where the standard vacuum energy for smooth parallel plates is recovered only in the limit $\mathcal{N}=1$ (or $\hat{f}=0$), consistent with the literature. This term constitutes the primary contribution, while the remaining terms account for the corrections arising from the interplay between temperature and surface geometry.

As previously shown, the resulting effective potential is finite after implementing the spectral $\zeta$-function framework and the renormalization conditions. Furthermore, we can determine the generation of a topological mass for the scalar field at one-loop order by evaluating at the stable minimum energy state $\Psi=0$. To this end, we apply the second condition from (\ref{Recond}) to the renormalized effective potential (\ref{VR}), yielding
\begin{eqnarray}\label{mTm}
    m_{T}^{2} &=& 
    m^{2}
    + \frac{g}{96a\left(a+\int_{0}^{1}\hat{f}(\beta)d\beta\right)}
    \int_{0}^{1}\mathcal{N}^{1/2}(\beta)d\beta
    \nonumber\\
    &+&
     \frac{g}{4\pi\xi\left(a+\int_{0}^{1}\hat{f}(\beta)d\beta\right)}
    \left(
    1
    - \frac{1}{\xi\left(m^{2} + \frac{\pi^{2}}{a^{2}}\int_{0}^{1}\mathcal{N}(\beta)d\beta\right)^{1/2}}
    \right)
    \nonumber\\
    &&
    \times
    \exp\left[-\xi\left(m^{2} + \frac{\pi^{2}}{a^{2}}\int_{0}^{1}\mathcal{N}(\beta)d\beta\right)^{1/2}\right].
\end{eqnarray}
In the limit $a\rightarrow\infty$, we observe that all terms cancel out, as expected. Consequently, the case $m=0$ can be evaluated without any impediment, then
\begin{eqnarray}\label{mT}
    m_{T}^{2} &=& 
    \frac{g}{\left(a+\int_{0}^{1}\hat{f}(\beta)d\beta\right)}
    \left\{
    \frac{1}{96a}\int_{0}^{1}\mathcal{N}^{1/2}(\beta)d\beta
    \right.
    \nonumber\\
    &&
    \left.
    + \frac{1}{4\pi\xi}
    \left(
    1
    - \frac{a}{\pi\xi\left(\int_{0}^{1}\mathcal{N}(\beta)d\beta\right)^{1/2}}
    \right)
    \exp\left[-\frac{\pi\xi}{a}\left(\int_{0}^{1}\mathcal{N}(\beta)d\beta\right)^{1/2}\right]
    \right\}
    \,.
    \nonumber\\
\end{eqnarray}
This result must be strictly positive, since it is evaluated in the minimum stable state. Let us examine each term to verify this condition. In the first factor, we note the presence of the roughness, which acts as a perturbation. This implies that the integral must be strictly positive, establishing once again that $g>0$. In the second term, the presence of temperature ensures the positivity of the expression within the parentheses for large values of $\xi$, regardless of the functional form of the roughness. Consequently, no instability issues arise in the result.

The entire analysis has been completed considering the low-temperature contribution, that is, for large but finite values of $\xi$. According to (\ref{mT}), we note that there are topological corrections to the mass generation due to the smooth roughness. In our results, it is always possible to perform a perturbative expansion of $\mathcal{N}$ around $\hat{f}$ to explicitly determine its contribution. On the other hand, to better visualize the result for the topological mass, we fix $\hat{f}=0$ (see (\ref{N})), thus, we obtain
\begin{eqnarray}
    m_{T}^{2} &=& 
    \frac{g}{96a^{2}}
    + \frac{g}{4\pi\xi a}
    \left(
    1
    - \frac{a}{\pi\xi}
    \right)
    e^{-\frac{\pi\xi}{a}}
    \,.
    \label{mTsin}
\end{eqnarray}
In this case, we see once again that there are no stability issues since $\xi\gg a$. The first term corresponds exactly to that obtained in \cite{Toms:1979ij} for the case of parallel plates.


\section{Conclusions}

In this work, we investigate the one-loop radiative corrections to the effective potential of a self-interacting scalar field theory in $3+1$ dimensions, subjected to nontrivial geometry configurations and in the low-temperature limit. These corrections are determined by geometric perturbations caused by roughness in the plates under Dirichlet boundary conditions. The analysis is restricted to a local adiabatic roughness regime, where the surface profile varies smoothly over length scales much larger than the plate separation. This smooth roughness is represented by a potential term that modifies the trivial structure of the eigenvalue spectrum (see Eq. (\ref{Delta})). Due to the complexity of the problem, we chose to use the WKB method to solve the spatial operator and, consequently, employed the contour integral method of the generalized $\zeta$-function to obtain a spectral function. Furthermore, we analyze the finite-temperature contribution in the low-temperature limit, determining the eigenvalues through perturbation theory. Under these conditions, our main objective was to study the Casimir energy density and the generation of a topological mass for the scalar field.

Once the effective potential is constructed, we apply a renormalization process by imposing the conditions (\ref{Recond}). Within to the spectral $\zeta$-function framework together with the applied renormalization process, and the adiabatic regime that restricts this problem to an analysis of smoothly varying surface deformations, we find that the resulting one-loop effective potential is finite and no additional counterterms are required. We emphasize that the present result should not be interpreted as an implication of the absence of ultraviolet counterterms, but rather as a consequence of the regularization methods and the adiabatic approximations applied in this model. 
Furthermore, we find that vacuum stability is achieved only with the trivial minimum-state solution $\Psi_0=0$, even in the presence of smooth roughness and temperature terms. The condition $V''_{\text{eff}}>0$ is guaranteed by the positivity of the coupling constant, $g>0$, as well as by the perturbative nature of the roughness and the rapid decay of the temperature-dependent contributions. Our central results are the Casimir energy density and the topological mass at one-loop, and their explicit dependence on the smooth roughness function and finite temperature, expressed in Eqs. (\ref{CasEn}) and (\ref{mT}). The thermal contribution was analyzed in the regime $\xi M\gg1$ and $\xi\gg a$, where the perturbative treatment of the lowest eigenvalues is justified. The presence of these geometric deviations modifies the known results by introducing corrections that shift the spectrum compared to the ideal case of flat plates. Furthermore, the explicit form for the Casimir energy density and the topological mass in this limit reveals that roughness introduces additional correction terms with different orders in power, while thermal effects become negligible due to the exponential decay of the corresponding terms. In the limit where roughness is negligible, i.e., $\mathcal{N}=1$ (see Eq. (\ref{mTsin})), our model accurately reproduces the well-known results reported in \cite{Toms:1979ij}, thus supporting the consistency of the proposed analysis.

Finally, the interaction between the field and the geometry of the plates provides a consistent description of both the stable vacuum energy and the topological mass generation. In this context, within the adiabatic regime considered here, smooth roughness corrections provide a nontrivial contribution to the Casimir energy density and topological mass generation. As is well known, these results highlight the crucial role of nontrivial geometries in determining the stability of the vacuum state, opening the possibility for further developments involving finite-temperature configurations, different boundary conditions, external fields, and more complex geometrical setups.


\end{document}